\documentclass{amsart}
\newtheorem{thm}{Theorem}[section]
\newtheorem{prop}[thm]{Proposition}
\newtheorem{lem}[thm]{Lemma}
\newtheorem{cor}[thm]{Corollary}
\theoremstyle{definition}

\usepackage[dvips]{graphicx}
\usepackage[usenames,dvipsnames]{color}
\usepackage{epsfig}
\usepackage{rotating}
\usepackage{amssymb}
\usepackage{amsmath,amsfonts}
\usepackage[boxed]{algorithm}
\usepackage{verbatim}

\def\0{{\mathbf 0}}

\newcommand{\F}{\mathbb{F}}
\newcommand{\Z}{\mathbb{Z}}

\begin{document}
\title{Lower bounds on the minimum distance of long codes in the Lee metric}
\author{ Hugues Randriam}
\address{Telecom ParisTech,46 rue Barrault, 75634
Paris Cedex 13, France.}
\email{randriam@enst.fr}
\author{ Lin Sok}
\address{Telecom ParisTech,46 rue Barrault, 75634
Paris Cedex 13, France.}
\email{sok@enst.fr}
\author{ Patrick Sol\'e}
\address{Telecom ParisTech, 46 rue Barrault, 75634
Paris Cedex 13, France.\\and\\ Math Dept of King Abdulaziz University, Jeddah, Saudi Arabia }
\email{sole@enst.fr}
\begin{abstract}
The Gilbert type bound for codes in the title is reviewed, both for small and large alphabets. 
Constructive lower bounds better than these existential bounds are derived from geometric codes, either over $\F_p$ or $\F_{p^2},$ or over even degree extensions of 
$\F_p.$ In the latter case the approach is concatenation with a good code for the Hamming metric as outer code and a short code for the Lee metric as an inner code.
 In the former case lower bounds on the minimum Lee distance are derived by algebraic
geometric arguments inspired by results of Wu, Kuijper, Udaya (2007).
\end{abstract}

\maketitle
\section{Introduction}
The Lee metric was introduced in the coding theoretic literature as a way to analyze codes over large alphabets applied to phase modulation \cite{B}. More recently, it was used successfully
 in the correction of errors in constrained memories \cite{RoS}. The main challenge in Lee codes has always been to generalize the known results
for the Hamming metric at the price of added technical complexity. For instance the analogues of the  classical bounds of Hamming and Gilbert that are derived in \cite{A} and are reviewed below
do not even admit a closed form. In the present work we will strive to generalize what is perhaps the most unexpected and most difficult result of algebraic coding theory in the last century: the Tsfasman-Vladut-Zink
bound \cite{TVZ} that shows the existence of families of codes strictly better than the Varshamov-Gilbert bound, a non-constructive existential bound.\\

We begin by reviewing the known analogues of the Gilbert bound in the Lee metric (we reserve the term Varshamov bound for algorithmic constructions of generator matrices).
 The analogue of the Gilbert bound was derived first for odd alphabet size by Astola \cite{A}, using Lagrange multipliers, and for all $q$ by Gardy-Sol\'e \cite{GS} using saddle point approximation. In the same paper a bound valid for so-called
large alphabets (say  $q\ge 2n+1$) was introduced. All these bounds are non-constructive; in the present note we compare them with constructive bounds obtained from
algebraic geometric constructions. In particular a bound on the Lee minimum distance of geometric codes over a prime field was derived recently in \cite{WKU}. Unfortunately the authors 
of this interesting paper did not take into account the case of constant functions on the curve, and did not consider asymptotics. We correct this small technical oversight
and complete their work by deriving asymptotic estimates of the parameters of the shortened codes. We also derive a variant of this bound over $\F_{p^2}$ that allows to use geometric codes on the TVZ bound when the Ihara function $A(q)$ is known exactly
$A(p^2)=p-1.$ In general estimating $A(p)$ for $p$ a prime is a difficult question \cite{MP}.
 This latter bound is compared with the natural idea of concatenating a good Hamming weight geometric code over an extension field with a BCH code over the base field
with a good Lee distance \cite{RoS}. Plotting graphs for asymptotic bounds
show that the corrected bound of \cite{WKU} is in general better than the concatenated code bound, which in turn outperforms the non-constructive Gilbert type bound.
\section{Definitions}
If $q$ is any natural number we recall that the {\bf Lee weight } of a symbol $x$ of $\Z_q$ is $\min(x, q-x).$ The Lee weight of a vector is then defined by adding the 
contribution of each symbol.
\\

 The geometric codes we will consider will be of the form $C(D,G)$ with $D,G$ disjoint divisors of an algebraic curve of genus $g$ over $\F_q.$ Here $D=P_1+\cdots+P_n,$
is a localizing set; a possible choice for $G$ is $rP,$ for some $P$ on the curve; and the codeword attached to a function $f$ in the $L(G)$ space is of the form
$$(f(P_1),\dots,f(P_n)).$$
We let $r=\deg(G),$ and assume throughout that $ r>2g-2$ so that the parameters of $C(D,G)$ are $[n,k]$ with $k=r-g+1.$
By the Ihara function $A(q)$ we shall mean the largest number of rational points of a curve over $\F_q$ per unit of genus, for genus going to $\infty.$ In symbols
$$A(q)=\limsup_{g \rightarrow +\infty}\frac{N_q(g)}{g},$$
where $N_q(g)$ is the largest number of rational points of a curve over $\F_q,$ also known as the Serre function. It is known that $A(q)\le \sqrt{q}-1,$ for all $q$ (Drinfeld/Vladut bound)
and by using modular curves or recursive towers that
$A(p^{2m})=p^m-1.$
The coding motivation to define the Ihara function is the so called TVZ bound. This asymptotic bound states

$$R+\delta\ge 1-\frac{1}{A(q)},$$
for $R$ the rate of the family of geometric codes considered and $\delta$ its relative distance.
\section{Gilbert bound}
Let $R(\delta)$ be the largest achievable rate of a family of codes of relative Lee distance $\delta.$
\subsection{Small alphabets}
The following bound was obtained in \cite[Theorem 2]{A}.
{\thm \label{astola} If $q=2s+1,$ then $R(\delta)\ge 1+\log_q \alpha\beta^{\delta s},$ where $\alpha,\beta$ are defined by
\begin{eqnarray*}
 \alpha+2\alpha \sum_{i=1}^s \beta^i&=&1,\\
\alpha \sum_{i=1}^s i\beta^i&=&\frac{\delta s}{2}
\end{eqnarray*}

 }
\subsection{Large alphabets}
 The following function plays the role of Shannon entropy in the present context.
$$L_q(x)=x\log_q x+ \log_q (x+\sqrt{x^2+1})-x \log_q (\sqrt{x^2+1}-1).$$ The following result was first derived in \cite{GS}.
{\thm If $q\ge 2\delta n+1$ and $\delta \le L^{-1}(1)\approx 0.37,$ then $$R(\delta)\ge 1-L(\delta).$$}
\section{Geometric codes}

\subsection{Concatenation}
By the TVZ bound \cite[Theorem 13.5.4]{HP} we know there are families of geometric codes over
$GF(Q)$ for $Q$ a square, with rate ${\mathcal R}$ and relative distance $\Delta$ satisfying
$${\mathcal R}+ \Delta \ge 1- \frac{1}{\sqrt{Q}-1}$$
We concatenate this geometric code with a code over $\Z_q$ of parameters $[n, k]$ and minimum
Lee distance $d_L$. We must assume therefore that $Q = q^k$. If $q$ is not a prime, we
label $\Z_q$ by the elements of $GF(q)$ in an arbitrary fashion. In order to apply the TVZ bound
we must assume $Q$ to be a square, or, equivalently $k$ to be even.

{\prop  With the above notation the rate $R$ and the relative Lee distance $\delta$ of the concatenated code satisfy 
$$\frac{R}{k/n}+\frac{\delta}{d_L/n}\ge 1-\frac{1}{q^{k/2}-1} . $$}

Using some BCH codes over $\Z_p$ as inner codes yields the following bound.

{\cor \label{concat} For each prime $p\ge 7$ and every integer $1\le t \le (p+1)/2,$ such that $p$ is congruent to $t+1 \mod{2},$ there is a family of Lee codes over $\Z_p$
with rate $R$ and relative Lee distance $\delta$ satisfying
$$\frac{R(p-1)}{p-1-t}+\frac{\delta(p-1)}{2t} \ge 1- \frac{1}{p^{(p-t-1)/2}-1} .$$}
\begin{proof}
For these inner codes $n=p-1,$ $d_L\ge 2t$ and $k\ge n-t,$ by \cite{RoS}.
\end{proof}

\subsection{Victorian bound}
In \cite{WKU} a Lee analogue of Goppa estimate was derived by three authors from Victoria state of Australia. There was a slight error in their bound but it can easily be fixed as follows:
{\thm Given an algebraic curve of genus $g$ over $\F_q$ having at least $n+1$ rational points, there are codes of parameters $[n-1,r-g]$ over $\F_q$ with Lee distance $$d_L\ge \frac{n^2-r^2}{4r},$$ for any integer $r$ in the range
$(2g-2,n).$}

\begin{proof}
The geometric codes $C(D,rP)$ of \cite{WKU} have non-constant codewords of Lee weight $\ge \frac{n^2-r^2}{4r},$ and constant codewords of Lee weight $n,2n,\dots$
To remove these constant codewords we shorten the code in an arbitrary position thus decreasing the dimension by one unit.
\end{proof}
The following Corollary is immediate.

{\cor \label{victoria} For a family of curves of genus $g\sim \gamma n,$ the rate $R$ of the attached family of codes of relative distance $\delta$ is 
$$R\ge -\gamma -2\delta+\sqrt{4\delta^2+1}.$$}
\begin{proof}
Assume that for $n\rightarrow \infty,$ we have $r\sim x n;$ then passing to the limit in $k=r-g,$ we get $x=R+\gamma.$ Using the above theorem we should have
$$P(x)=x^2+4\delta x-1\ge 0,$$
which happens for $x\ge 0$ only if $x$ is larger than the larger root of $P(x)=0.$
\end{proof}

The best choice of $\gamma$ is $\gamma=1/A(q)$, in particular if $q$
is a square one can take $\gamma=1/(\sqrt{q}-1)$ as in the TVZ
situation, but the resulting code may be non-linear over the ring $\Z_q.$
If $q=p$ is a prime, less is known about the $A(p)$, but for example from \cite{AM} we know that $A(5)\ge 0.727.$

\subsection{A new bound using descent of the base field}

Let $p$ be an odd prime.
Let $\{1,\alpha\}$ be a basis of ${\mathbb F}_{p^2}$ over ${\mathbb F}_p$,
so ${\mathbb F}_{p^2}={\mathbb F}_p\cdot 1+ {\mathbb F}_p\cdot\alpha \cong{\mathbb F}_p\times {\mathbb F}_p$. 
If $c\in({\mathbb F}_{p^2})^n$ is a word of length $n$ over ${\mathbb F}_{p^2}$, then using this identification
we get a word $\widetilde{c}\in({\mathbb F}_p)^{2n}$ of length $2n$, and likewise, a linear code $C$ of parameters $[n,k]$ over ${\mathbb F}_{p^2}$ gives
rise to a linear code $\widetilde{C}$ of parameters $[2n,2k]$ over ${\mathbb F}_p$.
We extend the definition of the Lee weight to ${\mathbb F}_{p^2}$ by setting the weight of a symbol $z=x+y\alpha\in{\mathbb F}_{p^2}$ (where $x,y\in{\mathbb F}_p$) as
$$wt_L(z)=wt_L(x)+wt_L(y),$$
so summing over coordinates we find
$$wt_L(\widetilde{c})=wt_L(c).$$
We also let $S(j)$ be the cardinality of the ``sphere'' of radius $j$ in ${\mathbb F}_{p^2}$, so
$$S(j)=|\{z\in{\mathbb F}_{p^2};\;wt_L(z)=j\}|$$
and $B(M)$ the cardinality of the ``ball'' of radius $M$,
$$B(M)=|\{z\in{\mathbb F}_{p^2};\;wt_L(z)\leq M\}|=\sum_{j=0}^M S(j).$$
Last, we let $W(t)$ be the sum of the weights of the $t$ smallest elements in ${\mathbb F}_{p^2}$, so formally
$$W(t)=\sum_{j=0}^M jS(j)+(M+1)(t-B(M))\quad\text{ for }B(M)\leq t\leq B(M+1).$$
In this formula we allow real values of $t$.

\medskip

\begin{figure}
	\centering{
		\includegraphics[width=12cm]{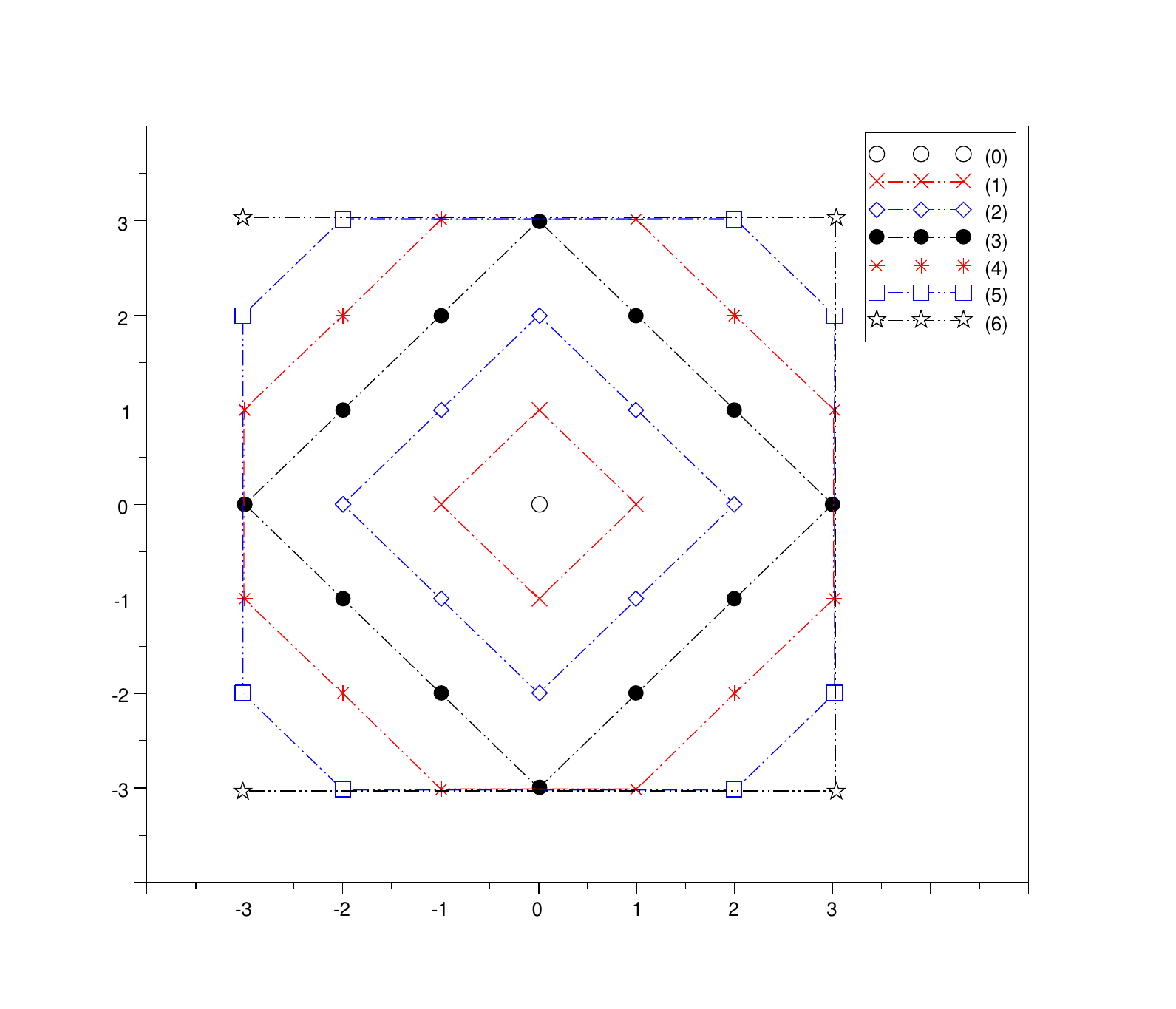}}
	\caption{\label{fig:leesphere} Lee spheres and Lee balls for ${\mathbb F}_{7^2}\cong{\mathbb F}_7\times {\mathbb F}_7$}
\end{figure}

\begin{figure}
	\centering{
\begin{tabular}{|c|c|c|c|c|c|c|c|}
\hline
$M$ & 0 & 1 & 2 & 3 & 4 & 5 & 6\\
\hline
$S(M)$ & 1 & 4 & 8 & 12 & 12 & 8 & 4\\
\hline
$B(M)$ & 1 & 5 & 13 & 25 & 37 & 45 & 49\\
\hline
$W(B(M))${\tiny ${}^{{}^{(1)}}$} & 0 & 4 & 20 & 56 & 104 & 144 & 168 \\
\hline
\end{tabular}

{\tiny $\qquad\qquad\qquad{}^{{}^{(1)}}$ $W$ is linear between $W(B(M)$ and $W(B(M+1))$}
}
	\caption{Values of the functions $S$, $B$, and $W$, for $p=7$}
\end{figure}

\medskip 

These quantities can be computed explicitly.
For $1\leq j\leq \frac{p-1}{2}$ the symbols $z\in{\mathbb F}_{p^2}$ having Lee weight $j$ are precisely the $z=\pm i\pm (j-i)\alpha$ for $0\leq i\leq j$,
and for $\frac{p+1}{2}\leq j\leq p-1$ these are the $z=\pm(\frac{p-1}{2}-i) \pm(j-\frac{p-1}{2}+i)\alpha$ for $0\leq i\leq p-1-j$, so
\begin{equation*}
S(j)=
\begin{cases}
1 & \text{ if } j=0\\
4j & \text{ if } 1\leq j\leq \frac{p-1}{2}\\
4(p-j) & \text{ if } \frac{p+1}{2}\leq j\leq p-1
\end{cases}
\end{equation*}
from which it follows
\begin{equation*}
B(M)=
\begin{cases}
1+2M(M\!+\!1) & \text{ if } 0\leq M\leq \frac{p-1}{2}\\
p^2-2(p\!-\!M)(p\!-\!1\!-\!M) & \text{ if } \frac{p+1}{2}\leq M\leq p-1
\end{cases}
\end{equation*}
and then, for $B(M)\leq t\leq B(M+1)$, a straightforward computation gives
\begin{equation*}
\begin{split}
W&(t)=\\
&\begin{cases}
\frac{2M(M+1)(2M+1)}{3}+(M\!+\!1)(t\!-\!B(M)) & \text{ if } M\leq \frac{p-1}{2}\\
\frac{p(p-1)(p+1)}{2}-\frac{2(p-\!M)(p-1-\!M)(p+1+2M)}{3}+(M\!+\!1)(t\!-\!B(M)) & \text{ if } M\geq\frac{p+1}{2}
\end{cases}
\end{split}
\end{equation*}
which can also be written
\begin{equation*}
\begin{split}
W&(t)=\\
&\begin{cases}
(M\!+\!1)(t\!-\!1)-\frac{2M(M+1)(M+2)}{3} & \text{ if } M\leq \frac{p-1}{2}\\
(M\!+\!1)(t\!-\!p^2)+\frac{p(p-1)(p+1)}{2}-\frac{2(p-\!M)(p-1-\!M)(p-2-\!M)}{3} & \text{ if } M\geq\frac{p+1}{2}.
\end{cases}
\end{split}
\end{equation*}
For example, we see that $W$ has maximal value
$$W(p^2)=W(B(p-1))=\sum_{i=-\frac{p-1}{2}}^{\frac{p-1}{2}}\sum_{j=-\frac{p-1}{2}}^{\frac{p-1}{2}}|i|+|j|=\frac{p(p-1)(p+1)}{2}.$$
Also, remark that in the case $M\leq\frac{p-1}{2}$, the expression does not depend on $p$.
And in the particular case $t=B(M)$ for $M\leq\frac{p-1}{2}$ we have $t=1+2M(M+1)$ so $M=\frac{1}{2}(\sqrt{2t-1}-1)$, so then $W(t)=\frac{2}{3}M(M+1)(2M+1)=W_{approx}(t)$
where
$$W_{approx}(t)=\frac{1}{3}(t-1)\sqrt{2t-1}.$$

\begin{figure}[ht]
	\centering{
		\includegraphics[width=12cm]{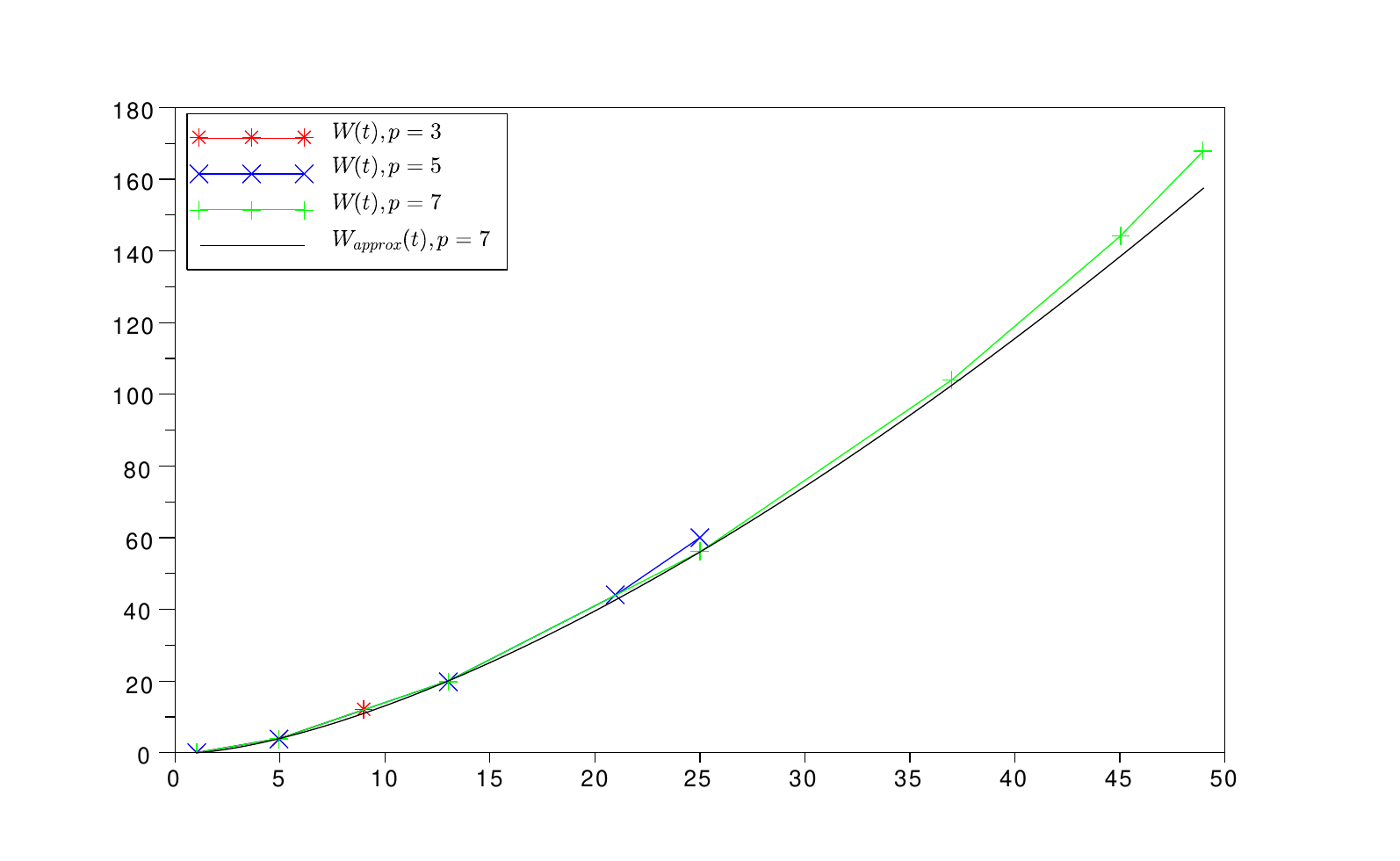}}
\caption{	\label{fig:w357} Graphs of $W_{approx}(t)$ and $W(t)$ for $p=3,5,7$}
\end{figure}


In fact one can show that $W_{approx}$ is a lower bound for $W$:

{\lem
\label{lem_lower_b}
We have
$$W(t)\geq W_{approx}(t)$$
for all real $t\geq1$.
}
\begin{proof}
First define $B_0(M)=1+2M(M+1)$ for all integer $M\geq1$, and then $W_0(t)=\frac{2}{3}M(M+1)(2M+1)\,+\,(M+1)(t-1-2M(M+1))$ for $B_0(M)\leq t\leq B_0(M+1)$.
By construction we have $B(M)=B_0(M)$ for $M\leq \frac{p-1}{2}$, and $B(M)\leq B_0(M)$ for $M>\frac{p-1}{2}$.
From this it follows $W(t)=W_0(t)$ for $1\leq t\leq\frac{p^2+4p-3}{2}$, and $W(t)\geq W_0(t)$ for $t>\frac{p^2+4p-3}{2}$.

Computing its second derivative, we see $W_{approx}(t)$ is a convex function of $t>1/2$. On the other hand $W_0(t)$ is a piecewise linear function,
and by construction both take the same values for all $t$ of the form $t=1+2M(M+1)$.
Since a convex function lies below its chords, we then conclude
$W(t)\geq W_0(t)\geq W_{approx}(t)$ for all real $t\geq1$.
\end{proof}

It is interesting to note that since $W_{approx}(t)$ is a convex interpolation of the piecewise linear function $W(t)$ in the range $1\leq t\leq\frac{p^2+1}{2}$, not only will it be a lower bound, but also a reasonably good approximation for it.
However, note that out of this range, while it is still a lower bound, the approximation can be slightly worse.
We compare the graphs of $W_{approx}(t)$ and $W(t)$  for some small primes in Figure \ref{fig:w357}.

{\thm \label{ls} Consider an algebraic curve of genus $g$ over $\F_{p^2}$ having at least $n+1$ rational points,
and an integer $r$ in the range $(2g-2,n)$.
Let $M$ be an integer such that $B(M)\leq n/r\leq B(M+1)$, e.g.
\begin{equation*}
M=
\begin{cases}
\left\lfloor \frac{1}{2}(-1+\sqrt{2n/r-1})\right\rfloor & \text{if $1\leq n/r\leq\frac{p^2+4p-3}{2}$}\\
\left\lfloor p-\frac{1}{2}(1+\sqrt{2p^2+1-2n/r})\right\rfloor & \text{if $\frac{p^2+4p-3}{2}<n/r\leq p^2$}.
\end{cases}
\end{equation*}
Then there are codes of parameters $[2(n-1),2(r-g)]$ over the prime field ${\mathbb F}_p$ with Lee distance 
 \begin{equation*}
\begin{split}
d_L&\geq rW(n/r)\\
&\qquad=\begin{cases}
 (M\!+\!1)n+\frac{(M+1)(2M^2+4M+3)}{3}\, r & \text{if $n/r\leq\frac{p^2+4p-3}{2}$}\\
  (M\!+\!1)n+\frac{2(M+1)(2M^2+4M-6pM-6p+3p^2)-p^3+p}{6}\, r & \text{if $n/r>\frac{p^2+4p-3}{2}$}\\
\end{cases}\\
&\geq rW_{approx}(n/r)=\frac{n-r}{3}\sqrt{\frac{2n-r}{r}}.
\end{split}
\end{equation*}
}

\begin{proof}
As before let $D=P_1+\cdots+P_n$ and $G=rP$ where $P,P_1,\dots,P_n$ are rational points on the curve, so that $C=C(D,G)$ has parameters $[n,r+1-g]$ over ${\mathbb F}_{p^2}$,
and $\widetilde{C}$ has parameters $[2n,2(r+1-g)]$ over ${\mathbb F}_p$.

Now following the idea from \cite{WKU} (already present in \cite{RoS}), we remark that if $c$ is a non-constant word in $C$, then any symbol from ${\mathbb F}_{p^2}$ can occur at most $r$ times in $c$.
This implies that we will have $wt_L(c)\geq wt_L(a)$ where $a=(a_1,a_2,\hdots,a_n)\in({\mathbb F}_{p^2})^n$ is constructed as follows:
\begin{itemize}
\item all symbols $z\in{\mathbb F}_{p^2}$ of Lee weight $0,1,\hdots,M$ occur in $a$ exactly $r$ times each
\item some symbols of Lee weight $M+1$ could occur in $a$, but not more than $r$ times each, and at least one of them less than $r$ times
\item no symbol of Lee weight greater than $M+1$ occur in $a$
\end{itemize}
where $M$ is such that $B(M)\leq n/r<B(M+1)$.
By construction we then have
$$wt_L(a)=rW(n/r).$$
To conclude we shorten $C$ to get rid of the constant codewords, decreasing the length and the dimension of $C$ by one, and the length and the dimension of $\widetilde{C}$ by two,
so this new $\widetilde{C}$ has parameters $[2n-2,2(r-g)]$ over ${\mathbb F}_p$ and Lee distance $d_L\geq rW(n/r)$,
as claimed.
\end{proof}

{\cor \label{victoria2} Suppose we are given a family of curves over ${\mathbb F}_{p^2}$ having at least $n+1$ rational points and genus $g\sim \gamma n$,
for some integers $n\to\infty$. Then from this, for any $R<1-\gamma$, we get a family of codes over ${\mathbb F}_p$ of asymptotic rate $R$ and asymptotic relative Lee distance 
\begin{equation*}
\begin{split}
\delta &\geq \frac{R+\gamma}{2}W(\frac{1}{R+\gamma})\\
&\geq \frac{1-(R+\gamma)}{6}\sqrt{\frac{2}{R+\gamma}-1}.
\end{split}
\end{equation*}
}
\begin{proof}
Apply Theorem \ref{ls} with $r\sim x n$ where $x=R+\gamma$ and pass to the limit.
\end{proof}

Remark we have $\gamma\geq\frac{1}{p-1}$, hence in the proof $n/r\sim 1/x\leq p-1\leq\frac{p^2+1}{2}\leq\frac{p^2+4p-3}{2}$, so,
of the two possible expressions for $W$ we only need the one corresponding to this range.
As already noted, this expression does not depend on $p$,
and moreover, this is the range where $W_{approx}$ is a good approximation of $W$.

The first inequality in this corollary can then be put in the form
\begin{equation*}
\delta\geq
\begin{cases}
\frac{1}{2}(1-(R+\gamma)) & \text{ if } 1\leq\frac{1}{R+\gamma}\leq 5\quad\text{ ($p\geq3$)}\\
\frac{1}{2}(2-6(R+\gamma)) & \text{ if } 5\leq\frac{1}{R+\gamma}\leq 13\quad\text{ ($p\geq5$)}\\
\frac{1}{2}(3-19(R+\gamma)) & \text{ if } 13\leq\frac{1}{R+\gamma}\leq 25\quad\text{ ($p\geq7$)}\\
\text{etc.} & \\
a_M-b_M(R+\gamma) & \text{ if } B(M)\leq \frac{1}{R+\gamma}\leq B(M+1)\quad\text{ ($p\geq 2M+3$)}\\
\end{cases}
\end{equation*}
where
$a_M=\frac{M+1}{2}$ and $b_M=\frac{(M+1)(2M^2+4M+3)}{6}$.
And the second inequality is a reasonably good approximation of the first and might be easier to handle in some applications, especially when $R+\gamma$ is small.

\medskip

We conclude with an equivalent formulation of the previous corollary.
Denote by
$$\alpha_{Lee,p}(\delta)$$
the largest real $R$ such that there exists a family of codes over ${\mathbb F}_p$ of asymptotic rate $R$ and asymptotic relative Lee distance $\delta$.
Introduce the function
$$f(x)=\frac{x}{2}W(\frac{1}{x})$$
and
$$g(x)=\frac{x}{2}W_{approx}(\frac{1}{x}).$$
Then:
{\cor \label{victoria3} Let $\gamma=\frac{1}{p-1}$. We have the lower bounds
\begin{equation*}
\begin{split}
\alpha_{Lee,p}(\delta)&\geq f^{-1}(\delta)-\gamma\\
&\geq g^{-1}(\delta)-\gamma.
\end{split}
\end{equation*}
}

These inverse functions $f^{-1}$ and $g^{-1}$ can be computed explicitly. This gives the expressions:
\begin{equation*}
f^{-1}(\delta)-\gamma=
\begin{cases}
1-2\delta-\gamma & \text{ if } 0\leq \delta\leq \frac{2}{5}\quad\text{ ($p\geq3$)}\\
\frac{1}{3}(1-\delta)-\gamma & \text{ if } \frac{2}{5}\leq \delta\leq \frac{10}{13}\quad\text{ ($p\geq5$)}\\
\frac{3}{19}(1-\frac{2\delta}{3})-\gamma & \text{ if } \frac{10}{13}\leq \delta\leq \frac{28}{25}\quad\text{ ($p\geq7$)}\\
\text{etc.} &  \\
c_M-d_M\delta-\gamma & \text{ if } C(M)\leq \delta\leq C(M+1)\quad\text{ ($p\geq 2M+3$)}
\end{cases}
\end{equation*}
where $c_M=\frac{3}{2M^2+4M+3}$,
$d_M=\frac{6}{(M+1)(2M^2+4M+3)}$,
$C(M)=\frac{M+1}{2}-\frac{(M+1)(2M^2+4M+3)}{6(1+2M(M+1))}$,
and
\begin{equation*}
g^{-1}(\delta)-\gamma=
\begin{cases}
 \left(\frac{-v-\sqrt{\Delta}}{2}\right)^{1/3}+\left(\frac{-v+\sqrt{\Delta}}{2}\right)^{1/3}+\frac{4}{3}-\gamma & \text{ if } \Delta \geq 0\\
 2\sqrt{\frac{-u}{3}}\cos\left(\frac{1}{3}\cos^{-1}(-\sqrt{\frac{27v^2}{-4u^3}})+\frac{2\pi}{3}\right) +\frac{4}{3}-\gamma &\text{ if } \Delta<0
\end{cases}
\end{equation*}
with $\Delta=6912\delta^6+2112\delta^4-\frac{16\delta^2}{3}$, $u=36\delta^2-\frac{1}{3}$ and $v=(48\delta^2-\frac{2}{7})$.

\section{Comparisons}
\subsection{Concatenation vs Victoria and descent}
\begin{figure}[ht]
	\centering{
		\includegraphics[width=12cm]{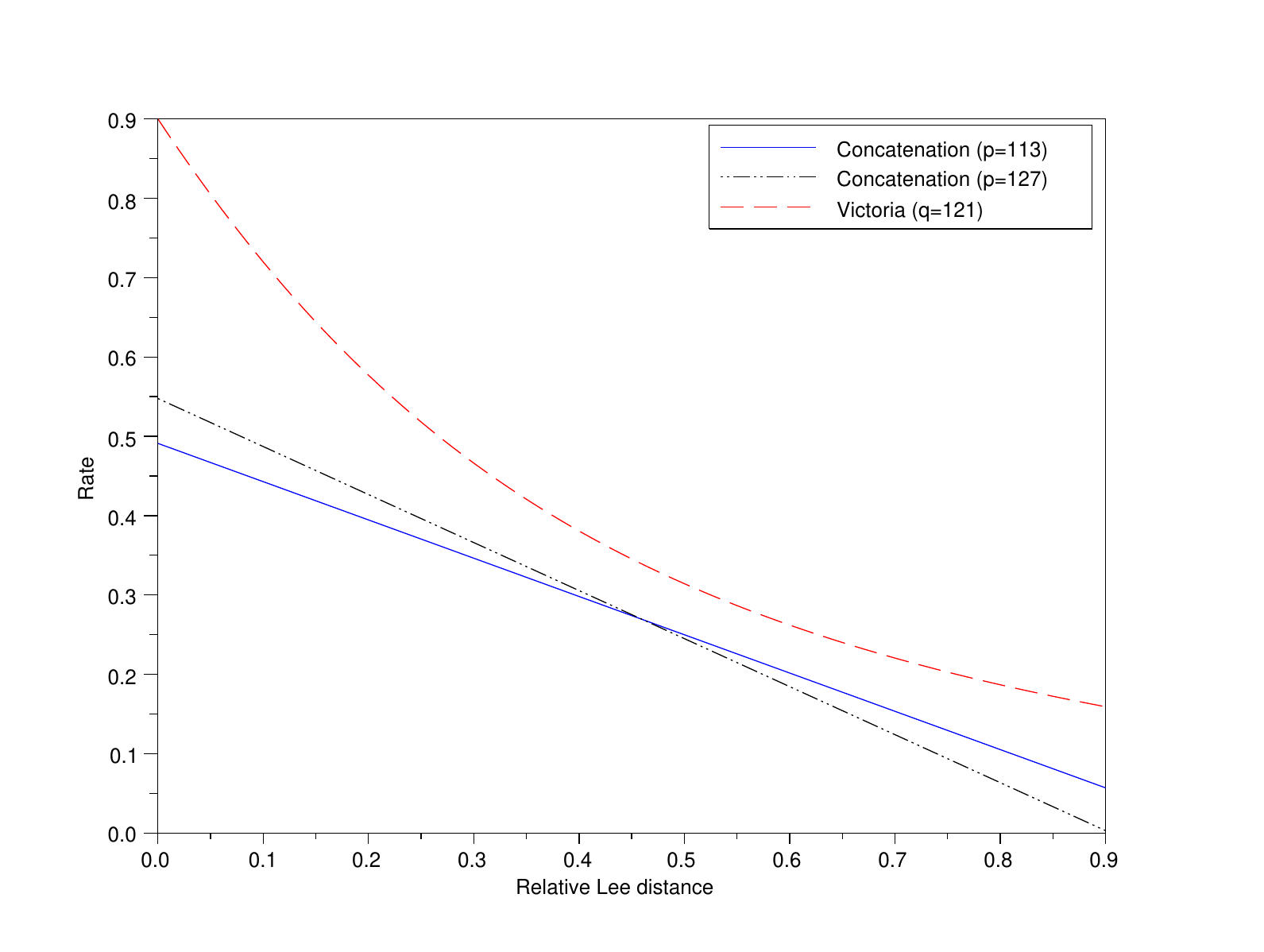}}
\caption{\label{fig:113}}
\end{figure}

\begin{figure}
	\centering{
		\includegraphics[width=12cm]{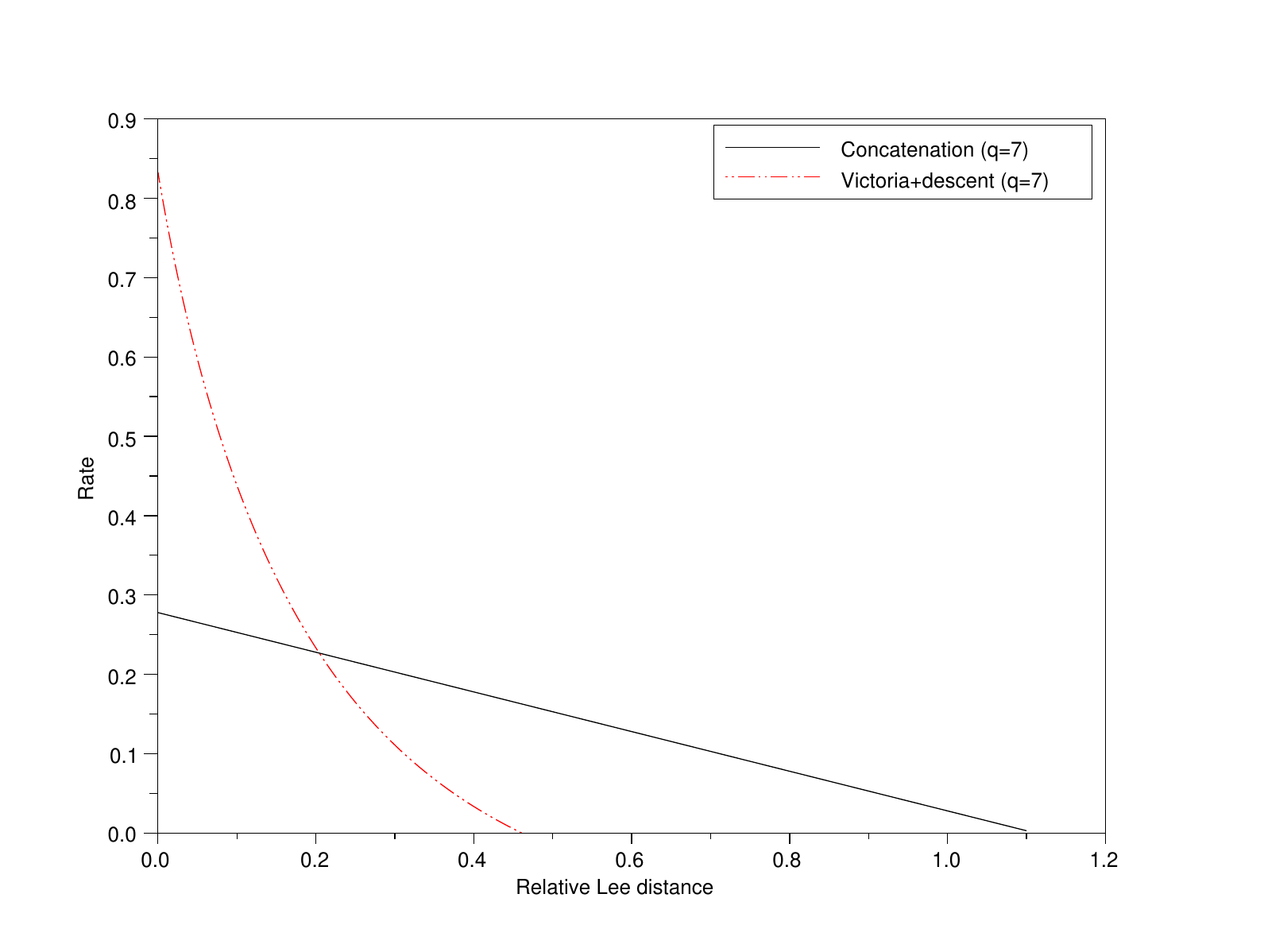}}
\caption{	\label{fig:concat.victo}}
\end{figure}

For large $p$ and $t$ of order $p/2,$  the bound of Corollary \ref{concat} can be approximated by
$$2R+\delta=1,$$
For large $q$, i.e $\gamma=0$, the bound  of Corollary \ref{victoria} can be approximated by
$$R\ge -2\delta+\sqrt{4\delta^2+1}.$$
Clearly the latter is above the former. \\
For finite values of $p$ in Corollary \ref{concat} and $q$ the square of a prime in Corollary \ref{victoria}, this first happens for $p=113$ or $127$ and $q=11^2(\approx p)$
as shown in Figure \ref{fig:113}.\\
For $q=p$ being prime, Corollary \ref{concat} is improved by Corollary \ref{victoria2}
where the case for $q=7$ is as shown in Figure \ref{fig:concat.victo}.

\subsection{Astola vs Victoria and descent}
\begin{figure}
	\centering{
		\includegraphics[width=12cm]{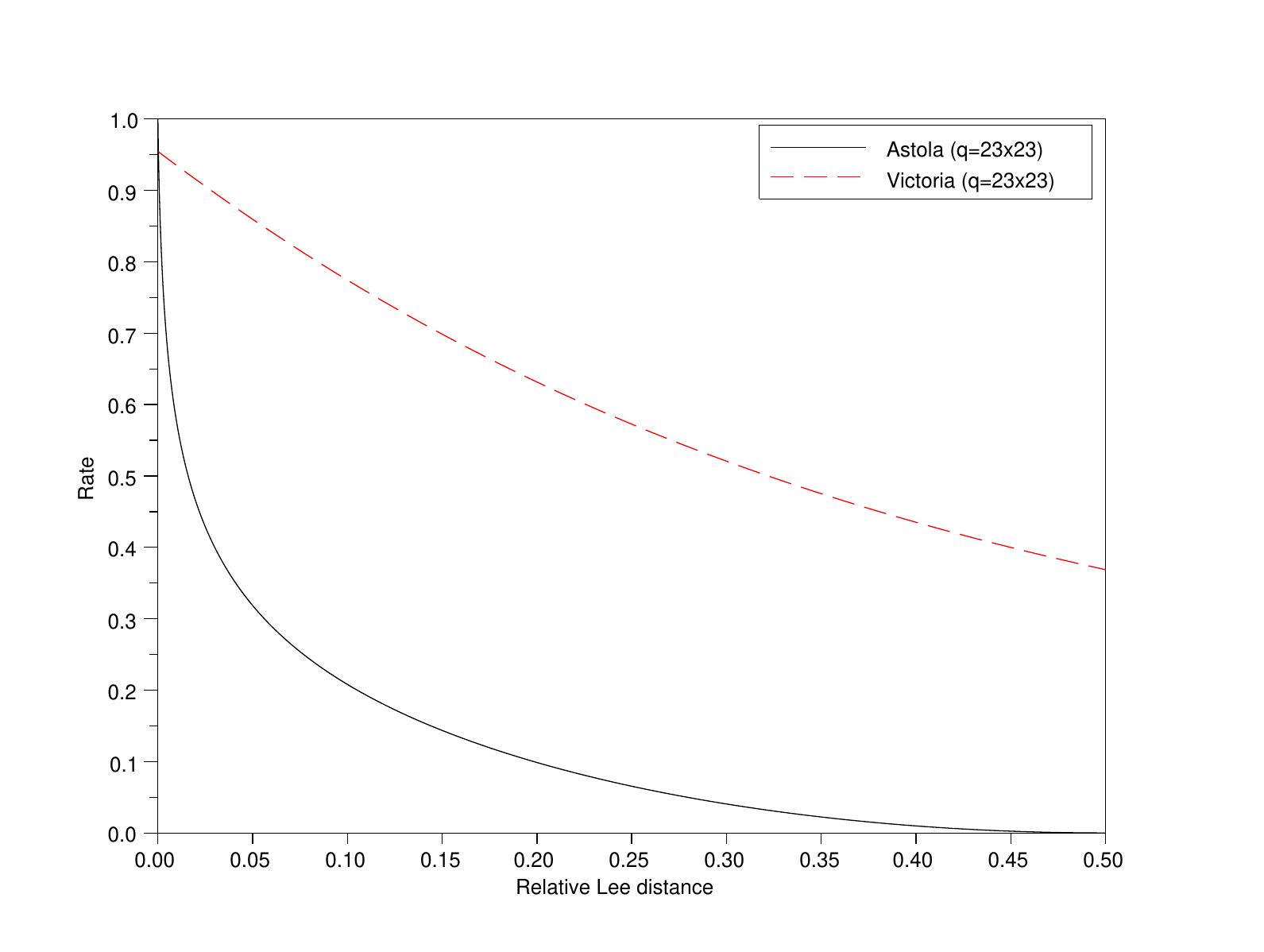}}
\caption{	\label{fig:23}}
\end{figure}

\begin{figure}
	\centering{
		\includegraphics[width=12cm]{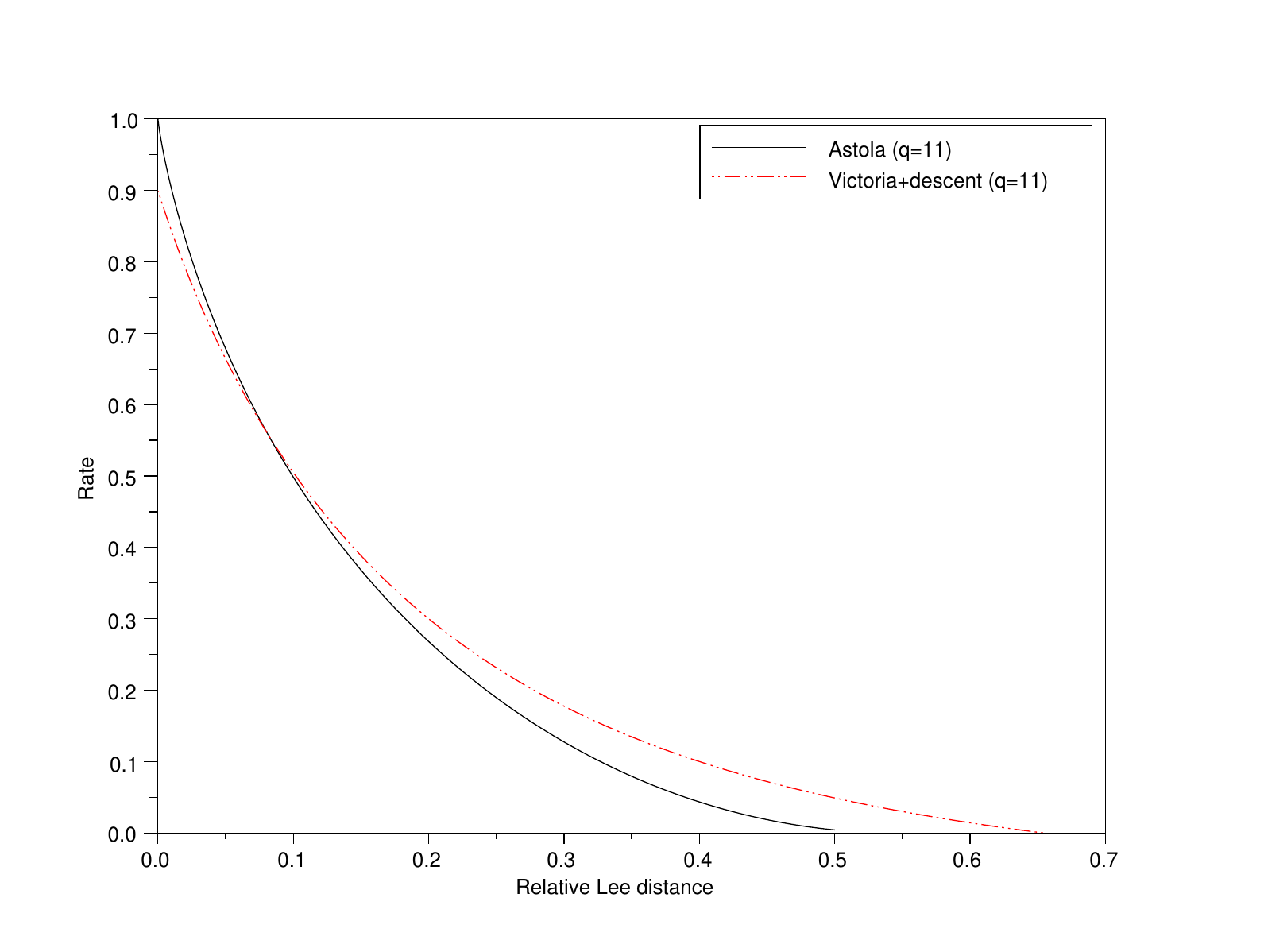}}
\caption{	\label{fig:asto.victo}}
\end{figure}

For $q$ at least $5^2$ Corollary \ref{victoria} is an improvement of Theorem \ref{astola} for $\delta\ge \delta_q=0.0308$. 
We list some small values of $q$ and such $\delta_q$ in Table \ref{delta_q} as well as give a graph of comparison for $q=23^2$ in Figure \ref{fig:23}. 
For $q$ being prime, Theorem \ref{astola} is improved by Corollary \ref{victoria2} where the case for the first prime $q=11$ is as shown in Figure \ref{fig:asto.victo}.

\begin{table}
\caption{\label{delta_q}}
\begin{center}
\begin{tabular}{|c|c|c|c|c|c|c|c|}
\hline
$q$& $5^2$ & $7^2$ & $11^2$ & $13^2$ & $17^2$ & $19^2$ &$23^2$\\
\hline
$\delta_q$ &0.0308&0.0095&0.0023&0.0014&0.0007&0.0005&0.0003\\
\hline
\end{tabular}
\end{center}
\end{table}

\pagebreak



\begin{thebibliography}{99}
\bibitem{AM} B. Angles, C. Maire
A note on tamely ramified towers of global function fields
Finite Field Appl. 8 (2002), 207--215.
\bibitem{A}J. Astola, On the asymptotic behaviour of Lee codes,Discr. Appl. Math {\bf 8} (1984)18--23.
\bibitem{B} E.R. Berlekamp,{\it Algebraic coding theory }, Aegean Park Press (1984).
 \bibitem{GS}D. Gardy, P. Sol\'e,``Saddle Point Techniques in Asymptotic Coding
Theory.''
Congr\`es Franco-Sovi\'etique de codage alg\'ebrique, Paris (1991),
Springer Lecture Notes in Computer Science 573 (1991) 75--81. {\tt ftp://ftp.cs.brown.edu/pub/.../91/cs91-29.pdf}
\bibitem{HP}Huffman W.C., Pless V.:{\it Fundamentals of Error-Correcting Codes.} Cambridge University Press, Cambridge
(2003).
\bibitem{MP} {\tt http://www.manypoints.org/}
\bibitem{RoS} R. M. Roth, P.H. Siegel, Lee-metric BCH codes and their application to constrained and partial-response channels. 
 IEEE Trans. Inform. Theory  40  (1994),  no. 4, 1083--1096.
\bibitem{TVZ} M.A. Tsfasman, S.G. Vladut, T. Zink, `` Modular curves, Shimura curves and codes better than the Varshamov-Gilbert bound,'' Math Nacrichen {\bf 109} (1982) 21--28.
\bibitem{WKU}X.-W. Wu, M. Kuijper and P. Udaya, Lower bound on minimum Lee distance of
algebraic–geometric codes over finite fields, Electronic Letters (2007) July, 19, Vol. 15, No 43.

\end{thebibliography}
\end{document}